\documentclass[sigconf,10pt,printfolios=true]{acmart}
\usepackage{booktabs} 
\usepackage{xcolor}

\usepackage{times}  
\usepackage{enumerate}

\usepackage{blindtext}
\usepackage[T1]{fontenc}
\usepackage[utf8]{inputenc}
\usepackage{newtxmath,bm,courier}
\usepackage{xspace,enumitem,fancyhdr,fancyref,wrapfig}
\usepackage{amsmath,amsfonts,bm}
\usepackage{tabularx,tablefootnote,wrapfig,hyphenat}
\usepackage[labelformat=simple,skip=0pt]{subcaption}
\usepackage{xfrac,sparklines}
\usepackage{algorithm,microtype}
\usepackage{algpseudocode}

\usepackage{tabularx}
\usepackage{wrapfig}
\usepackage{lipsum}
\usepackage{mathtools}
\usepackage{amsmath}
\usepackage{upgreek}
\usepackage{float}
\usepackage{hyperref,balance}
\usepackage{tabularx,multirow,booktabs,blindtext}

\newcommand{\parahead}[1]{\vspace*{1ex plus 0.25ex minus 0.25ex}\noindent %
  {\bfseries #1}}
\newcommand{\parabreak}{\vspace*{2ex plus 0.5ex}\noindent}

\newcommand{\Ku}{$\mathrm{K_u}$}
\newcommand{\shortname}{Wall-E}

\setitemize{itemsep=3pt,topsep=4pt,parsep=2pt,partopsep=0pt,leftmargin=1.5em}
\setenumerate{itemsep=3pt,topsep=4pt,parsep=1pt,partopsep=0pt,leftmargin=1.5em}

\begin{document}
\title{Towards Dual-band Reconfigurable Metamaterial Surfaces for Satellite Networking}


\author{Kun Woo Cho, Yasaman Ghasempour, Kyle Jamieson}
\affiliation{Princeton University}
\thanks{This material is based upon work supported by the National Science Foundation under Grant No.~RINGS-2148271}




\begin{abstract}

The first low earth orbit satellite networks for internet service
have recently been deployed and are growing in size, yet will face
deployment challenges in many practical circumstances of interest.
This paper explores how a dual-band, electronically tunable
smart surface can enable dynamic beam alignment between the satellite
and mobile users, make service possible in urban canyons, and 
improve service in rural areas.  Our design is the first of
its kind to target dual channels in the \Ku{} radio frequency band
with a novel dual Huygens resonator design that leverages
radio reciprocity to allow our surface to simultaneously steer and modulate
energy in the satellite uplink and downlink directions, and in 
both reflective and transmissive modes of operation.  Our surface,
\shortname{}, is designed and evaluated in an electromagnetic
simulator and demonstrates 94\% transmission efficiency and a 85\% reflection efficiency, with at most 6~dB power loss at steering angles over
a 150 degree field of view for both transmission and reflection.  
With $75cm^2$ surface, our link budget calculations
predict 4~dB and 24~dB improvement in the SNR of a 
link entering the window of a rural home in comparison to the free-space path and brick wall penetration, respectively.
\end{abstract}

\maketitle



\section{Introduction}
\label{s:intro}

\begin{figure}[t]
\begin{subfigure}[b]{.33\linewidth}
\includegraphics[width=1\linewidth]{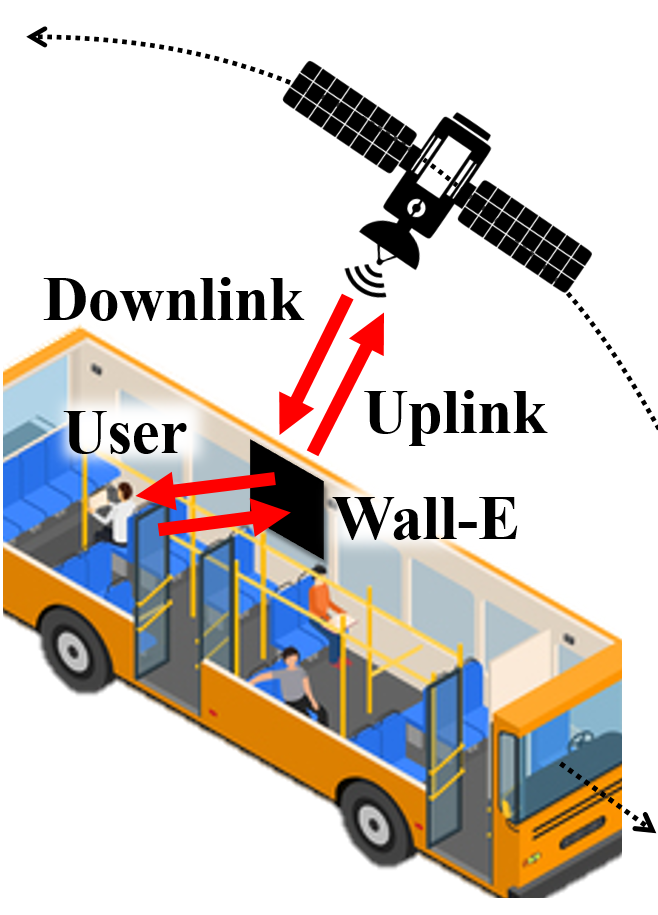}
\caption{Transportation}
\label{f:old_huygen}
\end{subfigure}
\begin{subfigure}[b]{.325\linewidth}
\includegraphics[width=1\linewidth]{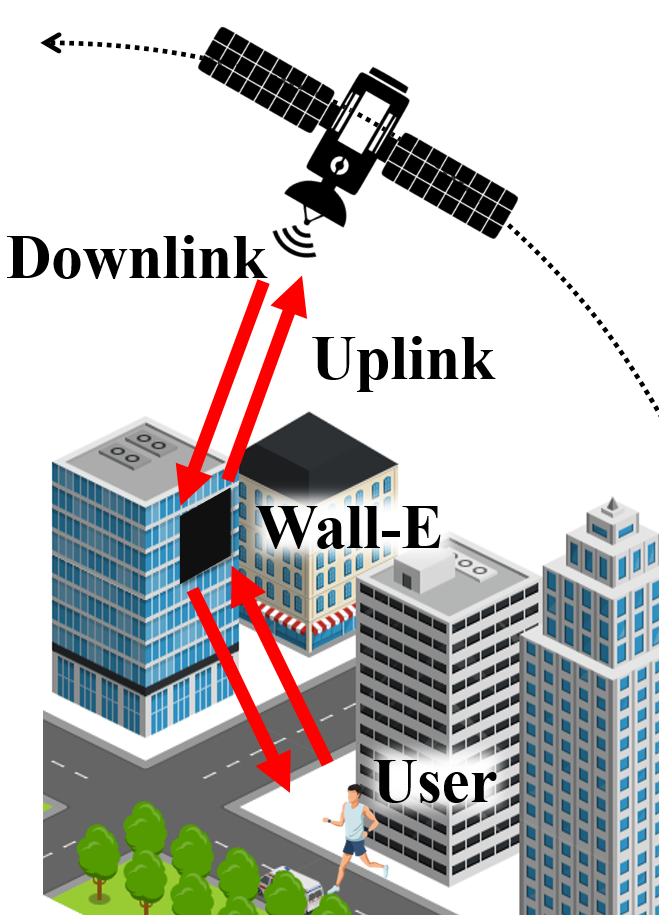}
\caption{Urban canyon}
\label{f:new_huygen}
\end{subfigure}
\begin{subfigure}[b]{.325\linewidth}
\includegraphics[width=1\linewidth]{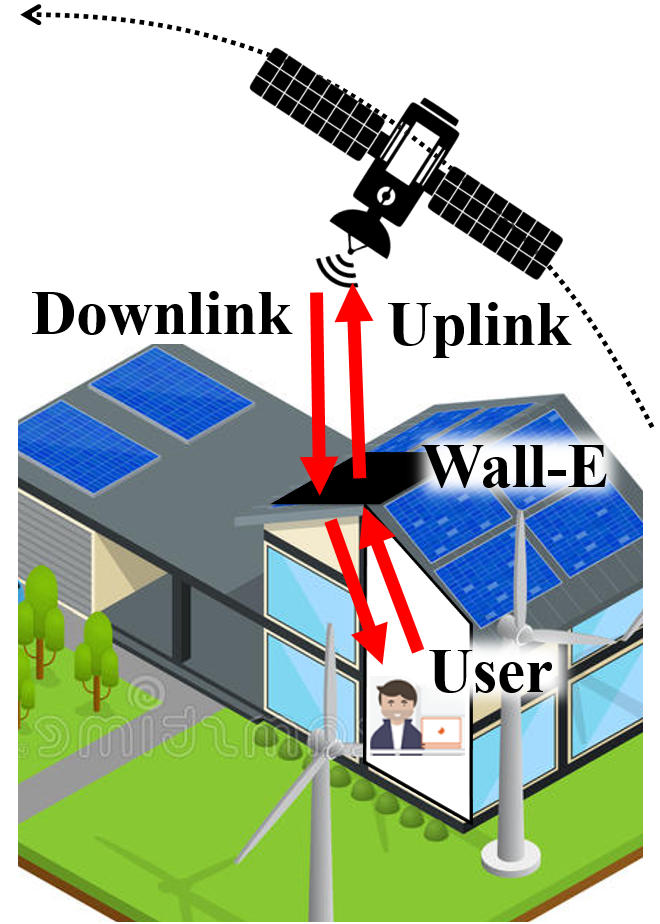}
\caption{Rural}
\label{f:analysis}
\end{subfigure}
\caption{Various use cases for a satellite smart surface.}
\label{f:designs}
\end{figure}

\begin{table}
    \centering
    \caption{Major current satellite internet service providers and 
    their primary frequency band allocations (GHz)~\cite{del2019technical}.}
    \label{tab:the_table}
\begin{tabular}{p{2.2cm}p{1.5cm}p{1.5cm}p{1.5cm}}
 \toprule
  & {\bf Starlink} & {\bf OneWeb} & {\bf TeleSat}\\
 \midrule
 Downlink   & 10.8--12.7   & 10.7--12.7 & 17.8--20.2\\
 Uplink     & 14.0--14.5   & 12.8--14.5 & 27.5--30 \\
 \bottomrule
\end{tabular}
\end{table}

Recently, there has been much interest in Low Earth Orbit (LEO) satellite data networking,
with multiple companies' networks in various deployment phases.  These networks consist
of constellations of hundreds of satellites that afford advantages 
in latency and coverage \cite{10.1145/3286062.3286075}: examples include
SpaceX's Starlink with a constellation of 4,425 satellites, 
and Telesat.
Current systems are generally designed with a dish antenna that the user mounts outside
the buildings requiring service, which communicates with the satellite in
both the uplink and downlink directions.
The dish antenna then communicates with the modem through a wire leading from
the dish into the building to a modem, which then wirelessly communicates with the user,
typically via Wi-Fi. 

While such networks are already deployed and seeing limited use,
we believe intelligent reconfigurable surfaces will expand their 
applicability and improve their performance in at least the following
three scenarios:

\parahead{1.~Rail\fshyp{}bus\fshyp{}airplane applications:} 
For best performance, transportation systems 
(in particular high speed rail and airplanes)
will demand adaptive systems to track the satellite currently 
serving the vehicle as well as handoff between satellites when the need
arises.  An electronically reconfigurable surface mounted on the windows
and\fshyp{}or skylights of the vehicles can enable 
dynamic beam alignment to users inside. 

\parahead{2.~Service in urban canyons:} Tall buildings in a city will
reduce satellite lines of sight and preclude areas of coverage at 
or near street level for satellite networks.  While 
5G\fshyp{}NextG wireless
coverage is maximized in cities, high frequency financial trading
gains an advantage by using such networks \cite{10.1145/3286062.3286075}
and so urban deployment remains relevant.  An electronically
reconfigurable surface mounted externally mid\hyp{}way up
a skyscraper can enable service at street level
via reflection off the building, while also allowing 
satellite signals to transmit into the building through 
the surface.

\begin{figure*}
\vspace{-5pt}
\includegraphics[width=1\linewidth]{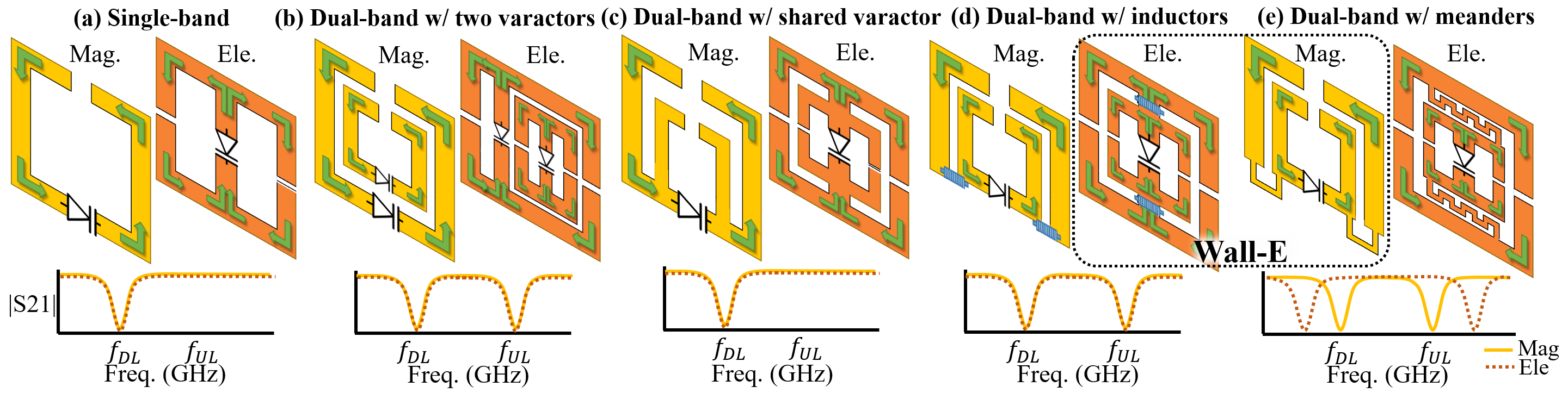}
\caption{The magnetic and electric meta-atom design considerations (\emph{top:} design schematics where the magnetic and electric meta-atom are colored in yellow and orange, respectively;  \emph{bottom:} transmission responses in magnitude $|S21|$). The electric and magnetic meta-atom inside a dotted black line are the designs selected for \shortname{}.}
\label{f:meta-atom}
\end{figure*}

\parahead{3.~Rural service:} While current LEO satellite systems 
require a dish and use a gateway to forward traffic between
the satellite link and client, an electronically reconfigurable
surface mounted on a window or skylight can refract the satellite link
into the home directly, getting rid of the outside dish.

The \Ku{} band (10.7--18~GHz) is a natural choice for such
low earth orbit satellite networks, as it has a longer
wavelength (25--17~mm) than the higher frequency
bands also in use (Table~\ref{tab:the_table}, mitigating the
impact of precipitation somewhat, yet also has a wavelength
short enough to create narrow beams for highly directional
communication to ground.  Since it has a short 
wavelength (25--17~mm) to experience loss 
when traversing heavy walls, it requires a line-of-sight (LoS) 
or near\hyp{}LoS (\emph{i.e.}, traversing only through a 
low\hyp{}loss material such as glass) path between the transmitter and receiver. 
A solution where instead of communicating with a dish relay through
a gateway, a nearby surface refracts or reflects
the satellite's signal satellite to the user could 
reduce ``outages'' due to transient blockage,
as it would allow path diversity, rerouting via the surface
to avoid blockage.
However, a key obstacle to realizing a smart surface design
is that the frequency duplex
division (FDD) communication in LEO satellite networks
complicates operation, because
such networks use different frequency sub\hyp{}bands 
in the uplink (upper \Ku{} band) and downlink (lower \Ku{} band) directions,
as Table~\ref{tab:the_table} shows. 


\parabreak{}This paper explores innovations in the design
space of LEO satellite networking with a Reconfigurable 
Intelligent Surface (RIS).  In the process, we
describe our early prototype surface design, \textbf{\shortname{}},
a dual\hyp{}band, metamaterial\hyp{}based 
RIS design. 

We first explore fundamental \Ku{}\hyp{}band RIS design.
In order to bring RIS-enhanced LEO networking to our scenarios, 
the surface should support both transmission (through the surface)
and reflection (off the surface) modes. 
Huygens' metasurfaces (HMSs) have shown to be promising in 
creating such \emph{transmissive} and reflective 
functionalities in practice~\cite{cho2021mmwall, ding2020metasurface, liu2018huygens, chen2017reconfigurable, zhang2018space}, thus achieving a full 
360 degrees of control over radiated
energy.  While the basic principles of Huygens unit cells
are known, designs that simultaneously make parsimonious use of electronic
components (varactors and inductors), resonate at two or more different 
frequencies (\emph{bi\hyp{}resonant}),
and achieve high efficiency are still open.

In LEO satellite networking, the process of aligning 
the physical wireless beam directions among user, surface, and satellite
be very complex as both the satellite and user moves.
Narrowing our design space to bi\hyp{}resonant Huygens RIS designs,
we next explore how to steer the uplink and downlink 
beams while preserving angular reciprocity, thus
speeding the process of the beam alignment for the uplink
via downlink transmissions, and vice\hyp{}versa.
This is of particular importance when both communication endpoints
are moving rapidly, which is the case in a transportation scenario,
satellite communicating with an airplane or train.
In such cases, the LEO satellite network's use of 
frequency duplexing division (FDD) allows for
real\hyp{}time, continuous feedback in both the uplink and
downlink directions facilitating the constant tracking 
of the endpoints with respect to the RIS, and associated continuous
updating of the RIS' steering angles.
    
Finally, we consider starting directions for RISs to enable full
end\hyp{}to\hyp{}end LEO network designs.  We consider the
handover process as the constellation of LEO satellites collectively
moves over the earth, necessitating a handoff 
from one satellite to another, serving each user.
The ability of the RIS to split uplink radio energy to two satellites
and simultaneously combine downlink radio energy from two satellites 
makes a \emph{soft} handover a possibility, which we explore further herein.

\section{Huygens Metamaterials}

By design, HMS\hyp{}based surfaces consist of a layer of co-located orthogonal electric and magnetic meta-atom, facing each other across dielectric substrate,
as shown in Fig.~\ref{f:meta-atom}(a). The key principle is that the pair of two meta-atoms introduces a discontinuity in the impinging electromagnetic field whereby the meta-atoms manipulate field attributes, including magnitude and phase. In order to achieve on-demand control of the reflective/transmissive pattern, we mount a tunable, voltage-controlled electric component, known as varactor, on each meta-atom. Since varactors draw only a couple-of-hundred microwatts order of power, \shortname{} consumes extremely low power. Unfortunately, HMS unit cells
resonate at only one frequency (\emph{mono\hyp{}resonanance}) 
\cite{PhysRevLett.110.197401} and thereby cannot 
act on the FDD links LEO satellite networks require. 

Leveraging existing mono resonant structures for satellite networking, two alternative solutions are possible. \textbf{Strawman~(i)---}building and deploying two single-band RISs (one for uplink and one for downlink). This approach would allow for FDD communication, but demands separate beam training for directional uplink and downlink, thereby doubling the overall delay of beam training.  This is an important process because the satellite trajectory is not fully deterministic---it is subject to turbulence and uneven gravitational forces \cite{najder2021quality,peng2018improving}---and the terrestrial user is often mobile. Hence, the required three\hyp{}party (LEO, RIS, user) beam training needs to be continuously performed for link maintenance. \textbf{Strawman~(ii)---}Partitioning the surface into two subsets, each resonating at a different frequency. This approach has the advantage of link reciprocity, \emph{i.e.}, since the downlink and uplink resonant elements are co-located, the optimum surface configuration for a downlink transmission is very close (if not exactly the same) as that of the uplink transmission. However, with such partitioning, the number of surface elements is reduced by a factor of two (given a fixed form factor) in each band. Hence, the reduced directivity gain might not be sufficient to close the long-range air-to-ground links. 

\section{Design}
\label{s:design}

We explore the key choices in our design space: we first discuss surface\hyp{}enhanced 
LEO networking that leverages exiting mono\hyp{}resonant structures and their 
shortcomings in realizing a directional, highly\hyp{}mobile link. Then, we explain 
our unique dual-band meta-atom design and illustrate its key properties in 
fast link establishment and mobility management.  




\subsection{Building the Surface: Meta-Atoms} 

We now explore novel directions in the design space of 
the Huygens unit cell, composed of a magnetic side and an electric 
side, which we discuss in turn. Figure~\ref{f:meta-atom}(a) 
illustrates a magnetic meta-atom structure that operates only in one frequency band.  
Here, the magnetic field of an incident electromagnetic wave induces a rotating current (denoted by green arrows) within the metallic loop (colored in yellow), which in turn produces its own magnetic field.
To manipulate the field response, the 
meta\hyp{}atom is integrated with a varactor diode, a voltage-dependent capacitor.
The magnetic meta\hyp{}atom is in essence a resonator 
consisting of both inductance and capacitance. Hence, the resonance response can be controlled via a varactor. 
Thus, a na\"{\i}ve approach to enabling bi-resonant unit cells would be to include two co\hyp{}located metal rings (the inner ring optimized for 
the higher uplink frequency of 15~GHz, and the outer optimized for the lower 
downlink frequency of 10~GHz), as shown in Fig.~\ref{f:meta-atom}(b). 
Although simple, this approach would require two separate varactors, increasing cost, insertion loss, and biasing complexity.

\begin{figure}
\begin{subfigure}[a.]{.49\linewidth}
\includegraphics[width=.89\linewidth]{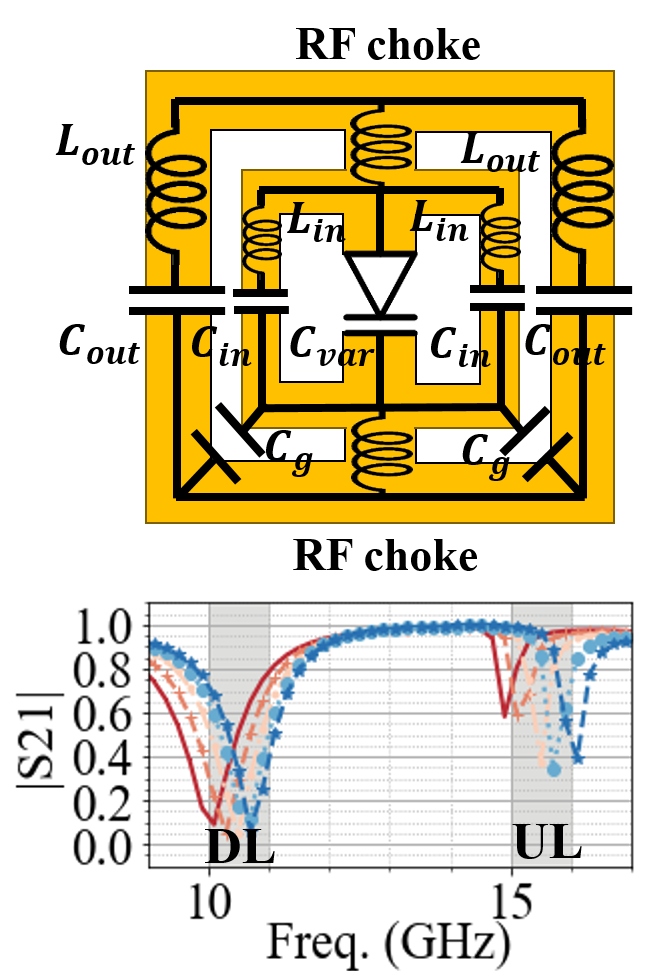}
\caption{Electric meta-atom}
\label{f:old_huygen}
\end{subfigure}
\begin{subfigure}[b.]{.49\linewidth}
\includegraphics[width=.88\linewidth]{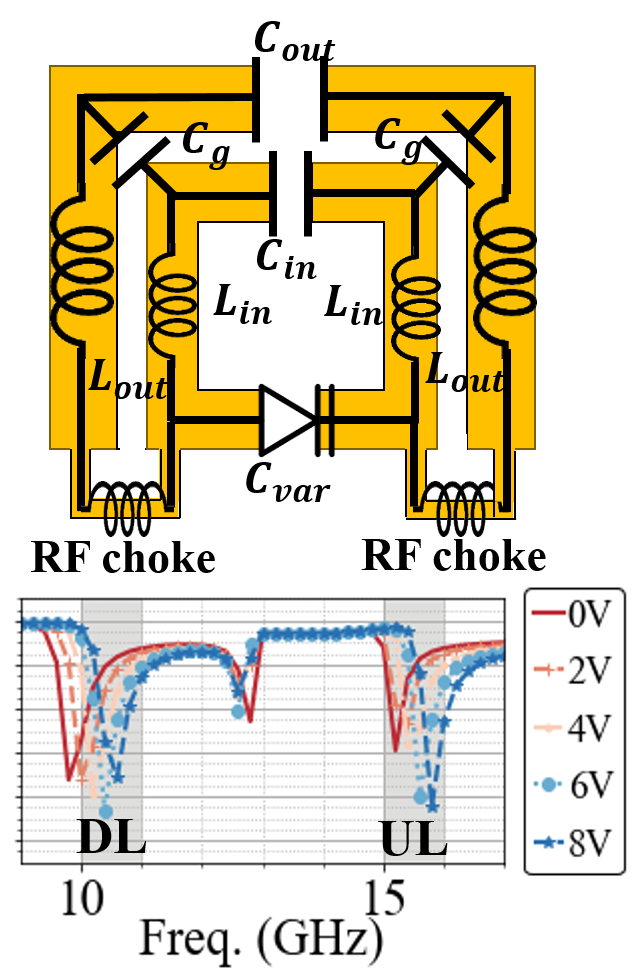}
\caption{Magnetic meta-atom}
\label{f:analysis}
\end{subfigure}
\caption{Equivalent circuit and its transmission response in magnitude across frequencies and voltages. The downlink and uplink frequency regions are colored in grey.}
\label{f:circuit}
\end{figure}

Instead, we want to control both outer and inner rings 
simultaneously, using a single shared varactor
(Fig.~\ref{f:meta-atom}(c)).
However, we find that with such a structure, only the outer
ring oscillates---thereby, the meta-atom effectively operates at
only a single frequency. 
In order to allow the passage of low frequency signals into the inner
ring, we load the connecting part with two RF chokes, 
which blocks the signal at higher frequency 
(for the inner ring) and passes the signal at lower frequency
(for the outer ring). 
We can design the choke in two ways: mount coil inductors 
(Fig.~\ref{f:meta-atom}(d)) or by bridging the outer and inner 
rings with a thin \emph{meander line} copper trace (Fig.~\ref{f:meta-atom}(e)).  
By adjusting meander width and length, we can apply 
a proper inductance value to choke off signals.
Since coil inductors increase insertion loss, we finally choose 
Fig.~\ref{f:meta-atom}(e) as our preferred magnetic side design candidate.


Figure~\ref{f:meta-atom}(a) also shows the electric side, 
resonating in one frequency band only. 
The electric field of an incident wave induces a rotating current 
within the metallic loop (colored in orange), which in turn produces 
its own electric field.
Similar to the magnetic meta-atom, Fig.~\ref{f:meta-atom}(c) shows two 
electric meta\hyp{}atoms with a shared varactor.
To properly control two rings using one varactor, 
we again connect two rings with RF chokes.
As shown in Fig.~\ref{f:meta-atom}(d), the outer ring has oscillating
currents at a lower frequency (downlink), and the inner
ring has its own oscillating current at a higher frequency (uplink). 
However, unlike the magnetic meta-atom, we do not use 
meander traces as a RF choke for the electric meta-atom,
because the meander trace would need to be placed in 
the gap between the two rings due to the different
structure of the electric meta-atom.
Increasing this gap, however, would create
a huge frequency difference between two rings as shown 
in the transmission response of Fig.~\ref{f:meta-atom}(e). Hence, 
we select the Fig.~\ref{f:meta-atom}(d) as our preferred 
electric side candidate.

\begin{figure}[t]
\includegraphics[width=.92\linewidth]{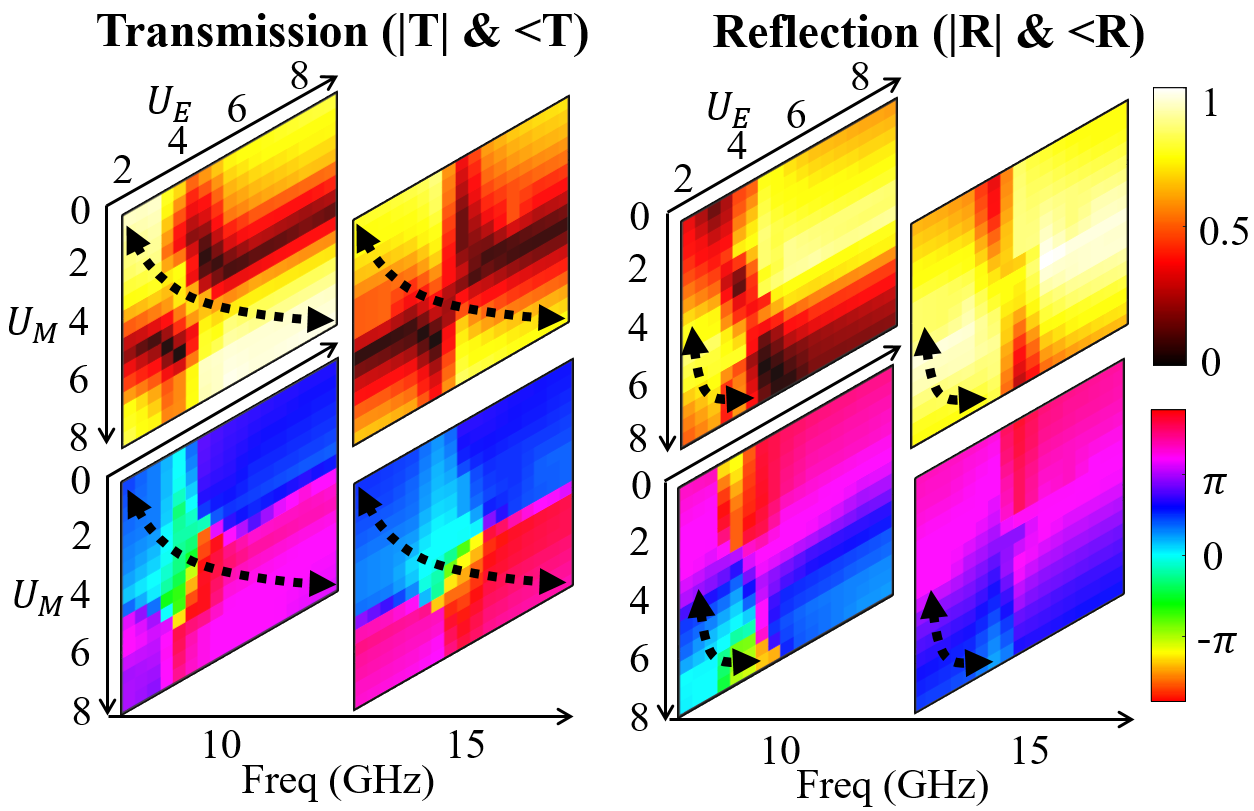}
\caption{Huygen's transmissive and reflective pattern in magnitude and phase at downlink and uplink frequency with different voltages applied to electric meta-atom $U_{E}$ and magnetic meta-atom $U_{M}$. The path denoted by the black dotted curve 
shows $360^{\circ}$, high amplitude phase coverage.}
\label{f:huygen_pattern}
\vspace{-10pt}
\end{figure}

\noindent
\textbf{Equivalent Circuit.}
Fig.~\ref{f:circuit} illustrates the candidate design's
equivalent circuit diagram, 
with the corresponding \emph{magnitude of transmission}
coefficient $|S21|$, across different frequencies, and
across different applied varactor control voltages. 
By definition, HMS currents oscillate at a resonant 
frequency $f=1/(2\pi \sqrt{LC})$ where $L$ is the
inductance and $C$ is the capacitance of the meta-atom.
In Fig.~\ref{f:circuit}, we see that each electric and magnetic 
meta-atom operates at two resonant frequencies, one at 
downlink and another at uplink. 
The resonant frequency for the downlink is largely
affected by the outer ring's inductance $L_{out}$ and capacitance $C_{out}$
while the inner ring's inductance $L_{in}$ and capacitance $C_{in}$ mainly 
determines the uplink's resonant frequency.
By increasing the voltage to the varactor, we 
decrease the total capacitance of the meta-atom, 
which, in turn, shifts the resonant frequencies, 
meaning that on each side, we can control both the outer and inner rings
with just a single varactor.

\begin{figure}
\includegraphics[width=.8\linewidth]{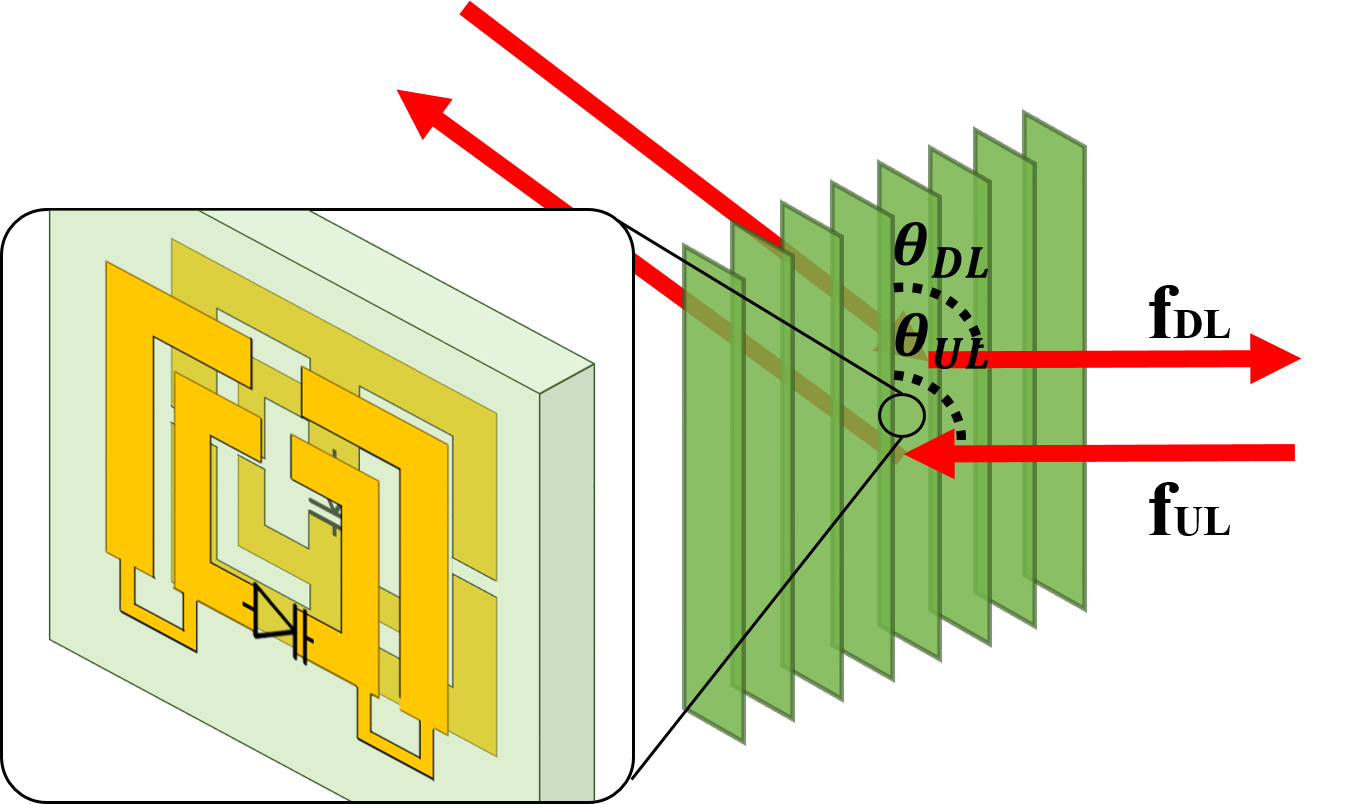}
\caption{\shortname{}'s bi-directional beam-steering in FDD communication. Due to its angular reciprocity, the steering angle of downlink $\theta_{DL}$ is equal to the steering angle of uplink $\theta_{UL}$.}
\vspace{-10pt}
\label{f:closeup}
\end{figure}

\noindent
\textbf{Huygen's Pattern.}
When we place the electric and magnetic meta-atoms together as shown in Fig.~\ref{f:closeup} and 
sweep the voltage across two varactors $U_{E}$ and $U_{M}$,
we obtain the transmission and reflection coefficient pattern, so called 
\emph{Huygen's pattern} as depicted in Fig.~\ref{f:huygen_pattern}.
This pattern demonstrates a full transmission and reflection phase
coverage of $360^{\circ}$ with near\hyp{}lossless
amplitude on the area marked by the black dotted curve.
While a single-band HMS obtains the Huygen's pattern at only one frequency, 
our design achieves this at \emph{both} frequencies,
enabling bi\hyp{}directional control of an FDD signal.
We will further optimize our inner rings for better uplink patterns.

\subsection{Establishing a Surface-Satellite Link}
Owing to the mobility of the LEO satellite as well as the end users, 
beam alignment plays a key role in maintaining the link quality of mobile satellite communication networks. 
We note that the coarse trajectory of the satellite is known a priori (and hence can be incorporated in beam adaptation protocols); yet, the exact real-time location of satellite cannot be perfectly predicted due to
the numerous factors like turbulence and uneven gravitational forces~\cite{najder2021quality, peng2018improving}. More importantly, the end point user is often mobile adding to the complexity of the three-party beam search between the satellite, user, and the surface. Conventional beam alignment protocols implement a trial-and-error scheme and test different potential directions sequentially. Extending such schemes to 
surface-enhanced satellite networks yield an increased delay as the beam training should be repeated at two different spectral bands.
In fact, \cite{guo2021two} demonstrates that a simulatenous uplink and downlink beamforming design in RIS-assisted FDD systems achieves more than $1.4$ times transmission rate over a one-way beamforming design
~\cite{yu2019miso, liu2020intelligent, hua2021reconfigurable}. 

On the other hand, \shortname{} can simultaneously steer the downlink and uplink beams at the same angle due to \textit{angular reciprocity}. Specifically, assume a certain biasing voltage configuration applied to the surface such that creates a transmissive steering angle of $\theta_{DL}$ for the incident downlink signal at 10 GHz, as shown in Fig.~\ref{f:closeup}. Due to the angle reciprocity, an uplink signal impinging the surface at $\theta_{UL}$ will be redirected toward the satellite location. Hence, angular reciprocity facilities fast beam alignments in FDD satellite networks as the surface configuration optimized for downlink transmissions works under the uplink communication and vice versa. 



\begin{figure}[t]
\begin{subfigure}[a.]{1\linewidth}
\centering
\includegraphics[width=.85\linewidth]{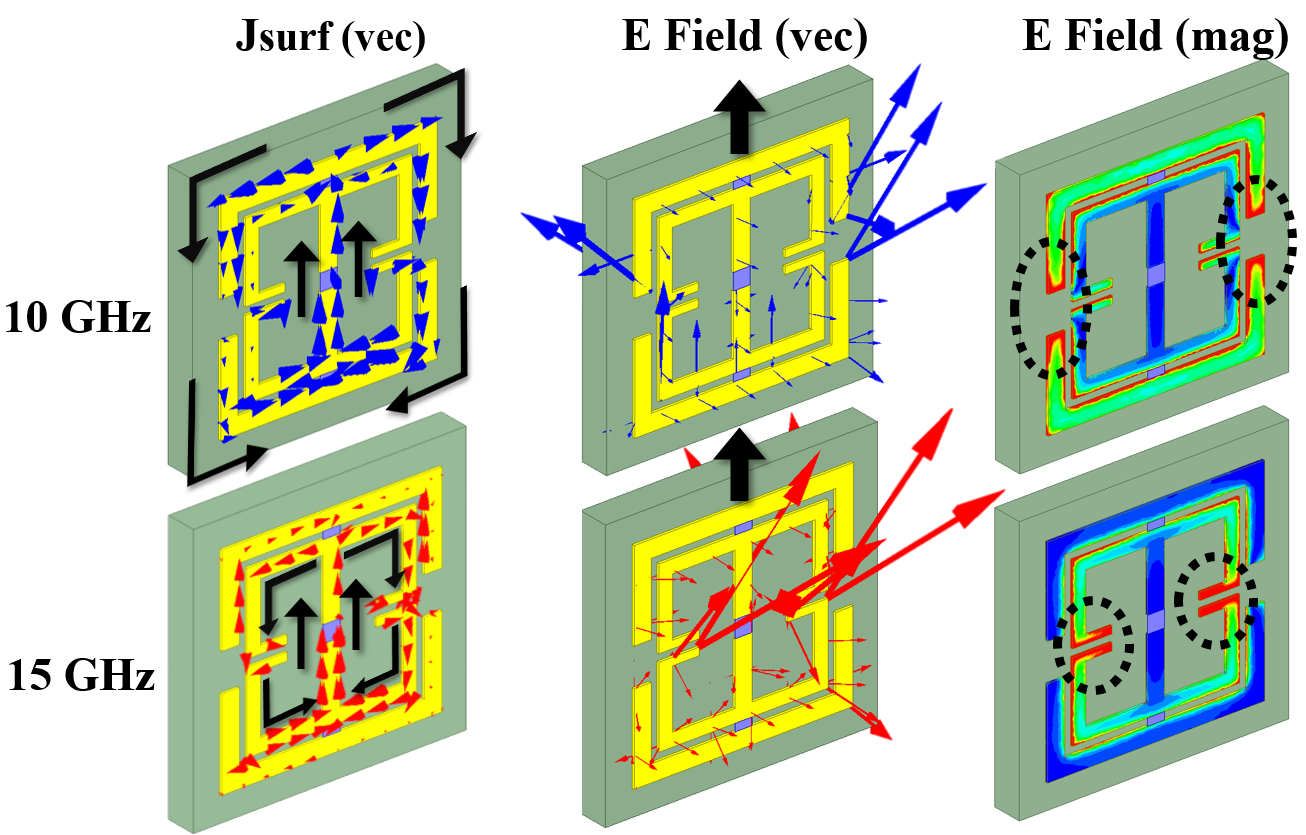}
\caption{Electric meta-atom.}
\label{f:electric_field}
\end{subfigure}
\begin{subfigure}[b.]{1\linewidth}
\centering
\includegraphics[width=.85\linewidth]{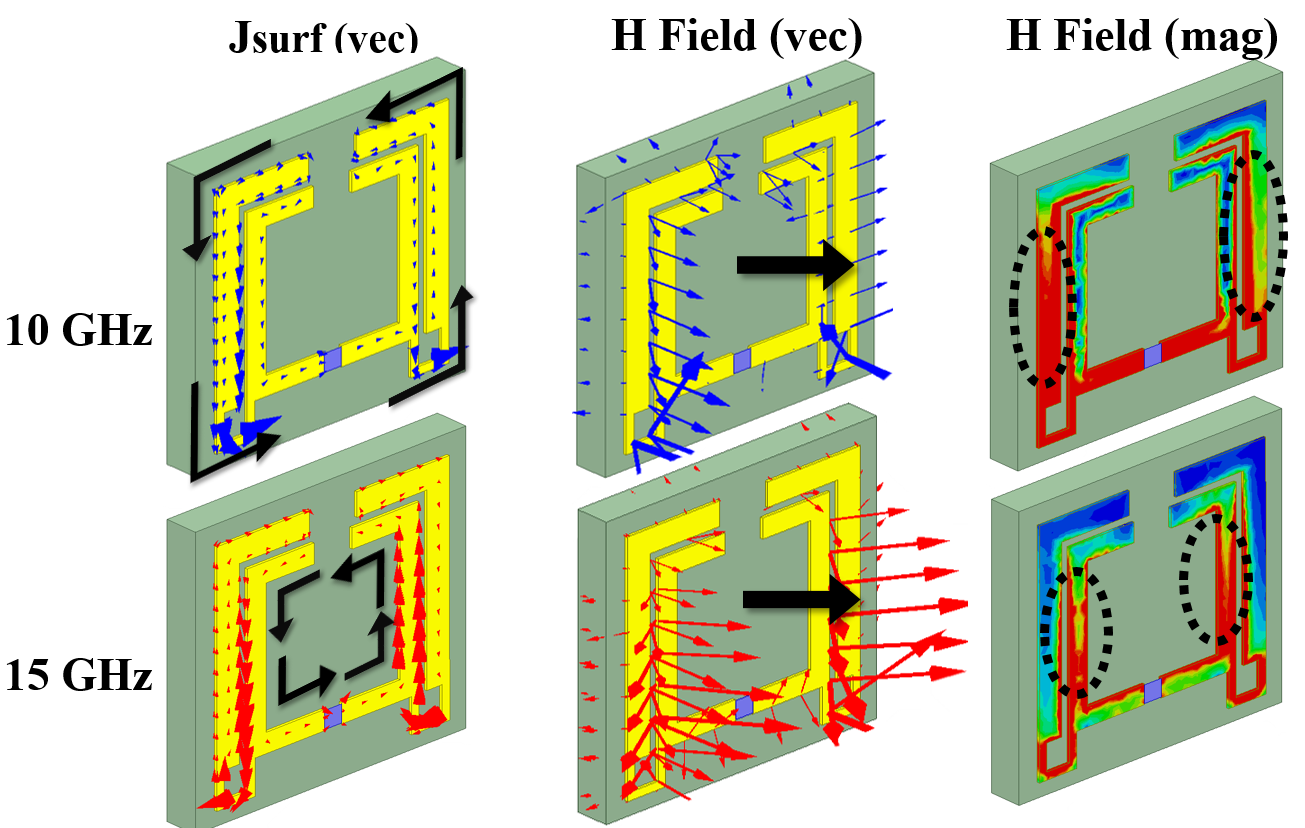}
\caption{Magnetic meta-atom.}
\label{f:magnetic_field}
\end{subfigure}
\caption{An electric meat-atom's surface currents and electric fields and a magnetic meta-atom's surface currents and magnetic fields at $10$ GHz and $15$ GHz.}
\label{f:fields}
\end{figure}

\subsection{Enhancing Satellite-Satellite Handover}
The fast movement of LEO satellites (around 7.5~km/s velocity relative to a reference point on the ground~\cite{cakaj2021parameters}) can cause multiple handovers resulting in an increase of RTT and a significant throughput drop. Even though other access networks (such as cellular networks) also experience handover, the impact of handovers on the transport layer and quality of service is relatively small, because of their relatively shorter RTT and thereby faster link recovery. We argue that an RIS-enhanced satellite network can substantially alleviate this problem. In particular, Wall-E supports soft handovers by allowing two (or multiple) satellites impinge on the surface at the same time. By carefully choosing the voltage configuration at each meta-atom, Wall-E achieves beam combining and steering. In this case, as the primary satellite fades away due to mobility, the secondary satellite will ensure a non-interrupted link. We highlight that such flexible handovers is owed to the on-demand wavefrom engineering at Wall-E.

\section{Feasibility}
\label{s:eval}

To project the feasibility of a RIS for LEO satellite networking, 
we simulate the performance of our design with HFSS EM simulation.
We also model our varactor based on its Simulation Program with Integrated
Circuit Emphasis (SPICE) model for an accurate capacitance value at our operating frequencies. In the future, we will fabricate and implement \shortname{}
and experiment with actual satellite signals. 

\noindent
\textbf{Near-Fields.} 
Figure~\ref{f:fields} illustrates the electric meta-atom's surface currents and electric fields (Fig.~\ref{f:electric_field}) along with the magnetic meta-atom's surface currents and magnetic fields (Fig.~\ref{f:magnetic_field}) at $10$ GHz and $15$ GHz. 
For both electric and magnetic meta-atom, the surface currents $J_{surf}$ oscillate on the outer ring at $10$ GHz while they oscillate on the inner ring at $15$ GHz. 
We denote the direction of $J_{surf}$ in black arrows, which conforms to Fig.~\ref{f:meta-atom}.
Similarly, the fields are excited by the outer ring in $10$ GHz by the inner ring in $15$ GHz. 
Fig.~\ref{f:fields} confirms the bi-resonate nature of the Wall-E meta-atoms.

\begin{figure}[t]
\begin{subfigure}[a.]{.48\linewidth}
\includegraphics[width=1\linewidth]{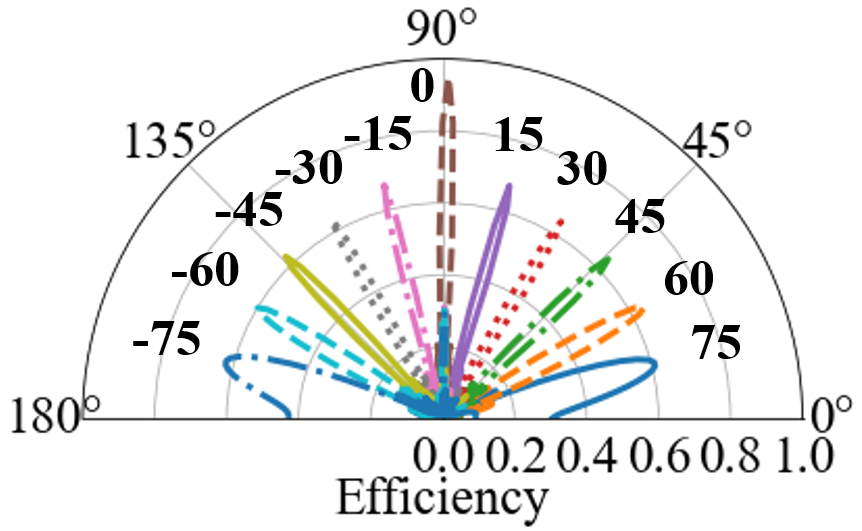}
\caption{Downlink transmission}
\label{f:down_trans}
\end{subfigure}
\begin{subfigure}[b.]{.48\linewidth}
\includegraphics[width=1\linewidth]{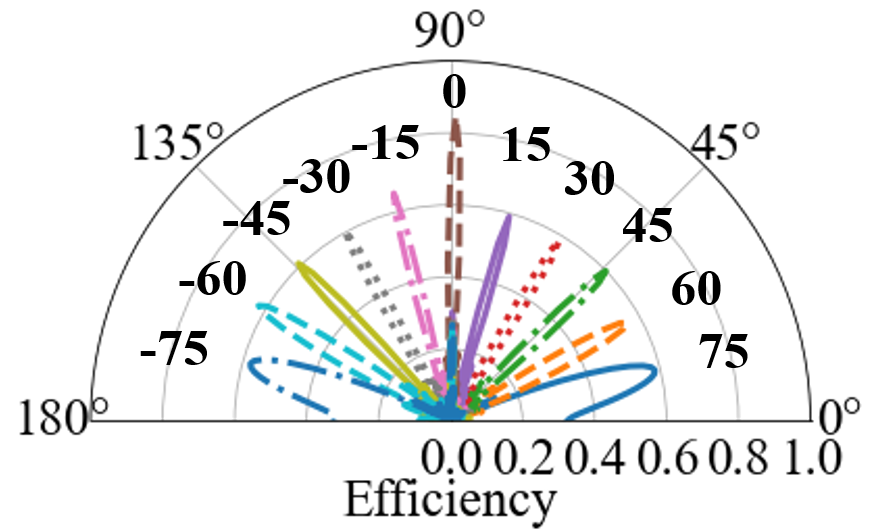}
\caption{Downlink reflection}
\label{f:down_refl}
\end{subfigure}
\begin{subfigure}[a.]{.48\linewidth}
\includegraphics[width=1\linewidth]{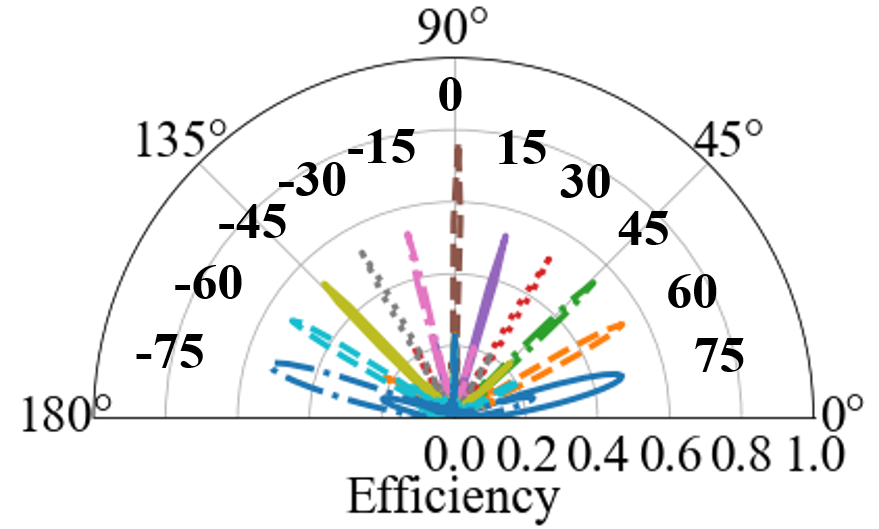}
\caption{Uplink transmission}
\label{f:up_trans}
\end{subfigure}
\begin{subfigure}[b.]{.48\linewidth}
\includegraphics[width=1\linewidth]{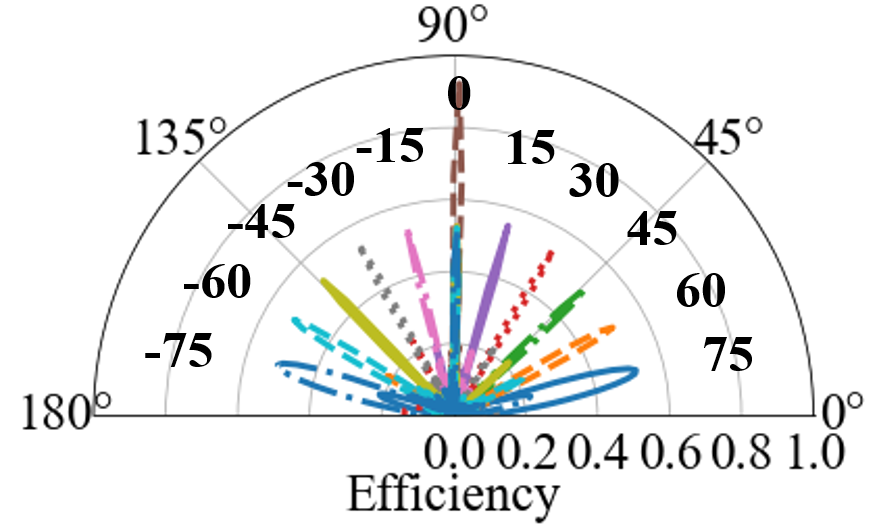}
\caption{Uplink reflection}
\label{f:up_refl}
\end{subfigure}
\caption{The transmission and reflection efficiency as \shortname{} steers the downlink and uplink beam.}
\label{f:beam_efficiency}
\vspace{-5pt}
\end{figure}

\begin{figure}[t]
\begin{subfigure}[a.]{.49\linewidth}
\includegraphics[width=1\linewidth]{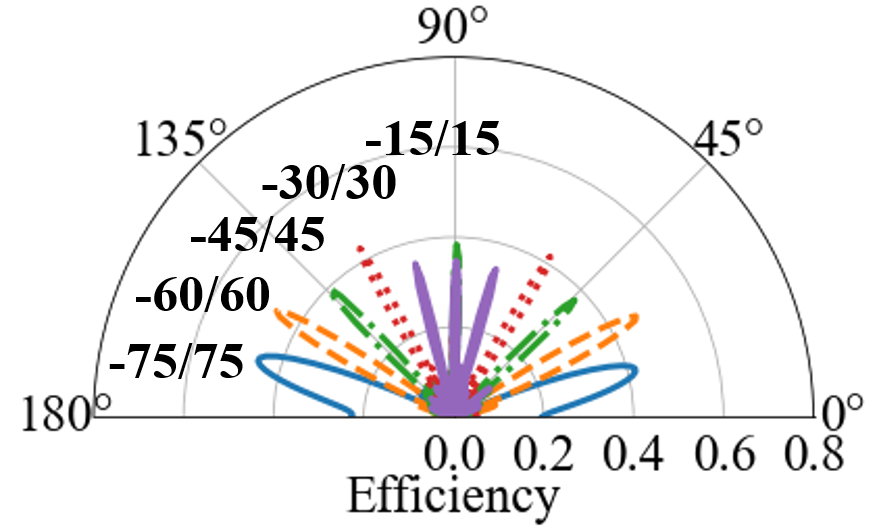}
\caption{Even beam split}
\label{f:down_split_even}
\end{subfigure}
\begin{subfigure}[b.]{.488\linewidth}
\includegraphics[width=1\linewidth]{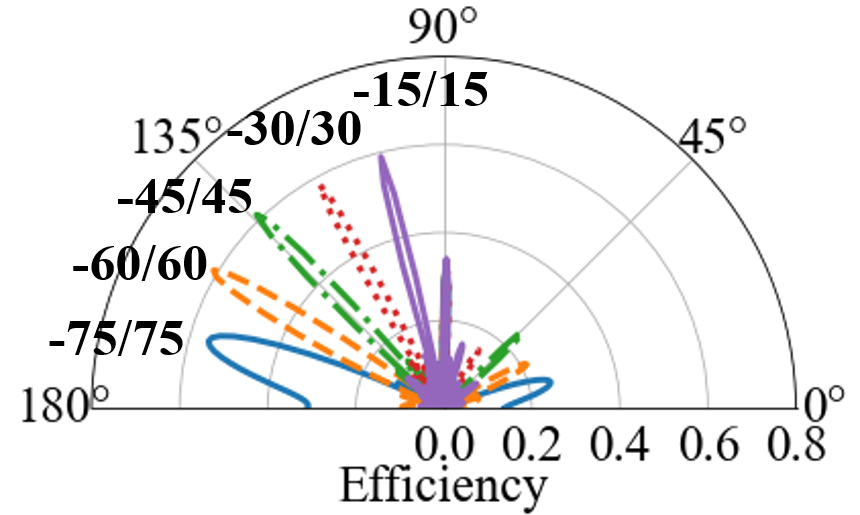}
\caption{Uneven beam split}
\label{f:down_split_uneven}
\end{subfigure}
\caption{The transmission efficiency of a beam splitted by \shortname{}. The power split is even for (a) and uneven for (b).}
\label{f:beam_split}
\end{figure}

\noindent
\textbf{Radiation Efficiency.}
In this section, we demonstrate a high efficiency of \shortname{}, which determines the end-throughput.
Specifically, we calculate the efficiency as a magnitude of an array factor at a desired angle:
\begin{equation}
\begin{array}{l}
AF {=} a_{0} {+} a_{1}e^{jkd(cos\theta)} {+} \dots {+} a_{N-1}e^{jk(N-1)d(cos\theta)}
\end{array}
\vspace{-2pt}
\label{eq:array_factor_hms}
\end{equation}
where $k=2\pi/\lambda$, $d$ is a meta-atom spacing, $\theta$ is a steering angle, and $a$ is a complex value chosen from Fig.~\ref{f:huygen_pattern}.
Fig.~\ref{f:beam_efficiency} shows the efficiency of \shortname{} as it steers the beam with the step of $15$-degree.
Specifically, Fig.~\ref{f:down_trans} demonstrates the efficiency of downlink transmission, which ranges from $62$\% to $94$\%. Similarly, Fig.~\ref{f:down_refl} illustrates the downlink efficiency as \shortname{} reflects the beam, which ranges from $60$\% to $85$\%. 
Fig.~\ref{f:up_trans} and Fig.~\ref{f:up_refl} reveal the efficiency of $50$\% to $80$\% for uplink transmission and reflection.
Moreover, we highlight \shortname{}'s beam-split performance in Fig.~\ref{f:beam_split} for a soft-handover. Here, we split the beam $150$, $120$, $90$, $60$, and $30$ degrees apart.
Specifically, the power is evenly divided for each beam on Fig.~\ref{f:down_split_even}, and it is unevenly splitted for Fig.~\ref{f:down_split_uneven} ($1/3$ on left and $2/3$ on right).
The result demonstrates that \shortname{} can tailor the beams in a flexible manner, which enables a highly-efficient relay and hand-over in FDD communication.

\noindent
\textbf{Link Budget.}
In this section, we analyze our back of the envelope calculation for closing a 1,150-km air-to-ground link. We formulate a link budget in decibel as follow:
\begin{equation} 
\vspace{-2pt}
    P_{rx} = P_{tx}+L_{d_{1}}+L_{window}+G_{\shortname{}, Rx}+L_{d_{2}}+G_{\shortname{}, Tx}+G_{rx}
\end{equation}
where $P_{tx}$ is a transmit power, including the transmitter gain. 
The Maximum Transmit EIRP of a most powerful satellite is $66.89$ dBW, which is equivalent to $97$ dBm~\cite{albulet2016spacex}.
We assume that the transmit power is $97$ dBm for downlink.
$L_{d_{1}}$ is a free-space path loss between the satellite and \shortname{}. 
Since an orbital height from Earth is approximately 1,150 kilometers~\cite{albulet2016spacex}, 
the free-space path loss $L_{d_{1}}$ is $-173.7$ dB for downlink and $-176.6$ dB for uplink.
$L_{window}$ is a $-4$ dB loss of window where the \shortname{} is placed.
Assuming $5$-meter distance between \shortname{} and user, 
$L_{d_{2}}$ is $-66.4$ dB. 
$G_{rx}$ is the receiving gain, which is equivalent to the gain of the user in downlink.
We assume that the gain of the user is $25$ dB.
Lastly, $G_{Wall-E, Rx}$ and $G_{Wall-E, Tx}$ is \shortname{}'s receiving and transmitting gain, respectively. Each is calculated based on the effective aperture, $A_{e}=\frac{\lambda^2}{4\pi}G$. 
Specifically, the surface gain $G_{Wall-E} = a_{\theta} 4\pi A_{e}/\lambda^2$ where $a_{\theta}$ is the radiation efficiency of \shortname{} at a steered angle $\theta$.
Finally, we obtain the SNR in decibel by subtracting the noise power from the signal power.

\begin{figure}[t]
\includegraphics[width=1\linewidth]{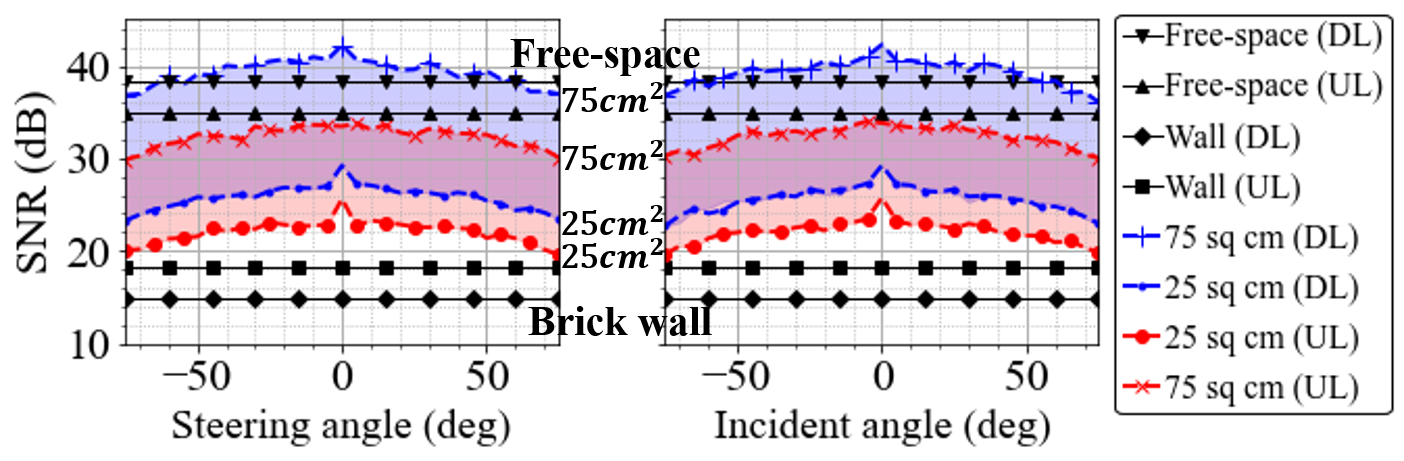}
\caption{Simulated SNR of transmissive links with varying surface size, steering and incident angle compared to the SNR of a free space path and wall penetration without \shortname{}.}
\label{f:snr}
\vspace{-5pt}
\end{figure}

Figure~\ref{f:snr} demonstrates the simulated SNR as \shortname{} steers the beam in two scenarios.
The first scenario has the incident beam perpendicular to the surface, and the surface steers from $-75$ to $75$ degrees. 
On the other hand, the second scenario varies the angle of the incident beam from $-75$ to $75$ degrees while \shortname{} steers the beam in a perpendicular direction.
For each scenario, we vary the size of \shortname{} and 
compare the simulated results against the free-space path and 
brick wall blockage in the absence of \shortname{}.
For both scenarios, the larger the surface is, the higher the SNR is. 
In particular, the SNR of $75cm^2$ sized \shortname{} is higher than 
the SNR of the free-space path for over a $100$ degree field of view.
Compared to the brick wall blockage scenario, 
$75cm^2$ sized \shortname{} provides approximately $24$ dB higher SNR.

\section{Related Works and Discussion}
\label{s:related}

\noindent

\noindent
\textbf{Dual-Band Metasurfaces.} 
Dual-band metasurfaces have recently gained attention for various applications however the existing architecture fall short in meeting at least one of our requirements, namely, on-demand and flexible reconfiguration, transmissive and reflective modes, and 360-degree coverage. In particualr,  
Han \textit{et al.}~\cite{han2021dual} introduced a dual-band transmission metasurface for S- and C-bands, which
provides wide-band operation with high transmission efficiency. However, the surface lacks dynamic configuration which makes it inapplicable in our highly mobile air-to-ground application. 
Saifullah \textit{et al.}~\cite{saifullah2021dual} proposed 
a dual-band programmable reflective metasurface that operates in C- and Ku- bands using PIN diodes. However, this design supports only the reflection mode; thereby, it's not suitable for through-wall applications. Further, the use of PIN diodes limits the phase shifting resolution and hence the beam steering efficiency. 
On the contrary, 
Rotshild \textit{et al.}~\cite{rotshild2021ultra} employs varactors to achieve a continuous  phase control. Unfortunately, this design is also reflection-only and is limited in angular coverage. To the best of our knowledge, \shortname{} is the first design prototype of a dual-frequency reflective/transmissive reconfigurable metasurface with a $360^\circ$ phase coverage and high radiation efficiency.

\textbf{Reflectarray Satellite Antennas.}
Prior works~\cite{encinar2006dual, encinar2011transmit, prado2015design, imaz2020reflectarray} have proposed the use of reflectarray antennas for space communication, where the reflectarray is placed on the satellite and is excited via the feed horn. Such an architecture can realize flexible steering as the reflected signal can be dynamically steered according to the array configuration.
Furthermore, ~\cite{patyuchenko2014digital, martinez2020multibeam} explored
the multibeam reflectarrays for the multispot coverage from the satellite.
Metasurfaces and reflectarrays are both spatially-fed structure composed of small elements. However, in contrast to existing efforts, we propose using a metasurface as an intermediate node (hence not co-located with the satellite nor end user) to increase path diversity.
In doing so, \shortname{}'s capability to support both transmission and reflection plays a key role.


\textbf{Wall-E for Satellite Link Aggregation.}
While gaining a lot of attention, the bandwidth of a single satellite path is unlikely to provide low-latency links comparable with a fiber path~\cite{10.1145/3286062.3286075}. However, aggregating abundant paths from many satellites within coverage zone of a user can, in principle, offer lower latency than a fiber path. We highlight that \shortname{} can play a crucial role in realizing satellite link aggregation by combining and steering the incident signals from multiple satellites into a desired direction. In the future, we will investigate novel scheduling algorithms for optimum coordination between multiple parties (all nearby LEOs, RIS, and user) and extend the prior efforts~\cite{vasisht2021l2d2, vazquez2018precoding, fan2021efficient} on link satellite scheduling that do not address multi-satellite and RIS-enhanced networking.

\clearpage
\balance
\bibliographystyle{abbrv} 
\bibliography{paper}
\clearpage

\end{document}